\documentclass[prb,aps,twocolumn,floats,showpacs]{revtex4}
\usepackage{epsfig}
\usepackage{graphicx}

\newcommand{\bfig}{\begin{figure}}
\newcommand{\efig}{\end{figure}}
\newcommand{\be}{\begin{equation}}
\newcommand{\ee}{\end{equation}}
\newcommand{\bea}{\begin{eqnarray}}
\newcommand{\eea}{\end{eqnarray}}

\begin{document}

\title{Dissipation-induced quantum phase transition in a quantum box}

\author{L\'aszl\'o Borda$^{1,2}$, Gergely Zar{\'a}nd$^1$ and  Pascal Simon$^3$}
\affiliation{
$^1$
Theoretical Physics Department,  Institute
of Physics, Budapest University of Technology and Economics, Budafoki
\'ut 8, H-1521\\
$^2$Research Group of Hungarian Academy of Sciences, Budafoki
\'ut 8, Budapest, H-1521\\
$^3$Laboratoire de Physique et Mod\'elisation des Milieux Condens\'es, CNRS et Universit\'e Joseph Fourier, 38042 Grenoble, France  
}
\date{\today}

\begin{abstract}
In a recent work, Le Hur has shown that dissipative coupling to gate 
electrodes may play an important role in a quantum box near its degeneracy point
[K. Le Hur, Phys. Rev. Lett. {\bf 92}, 196804 (2004)]: While quantum fluctuations 
of the charge of the dot tend to round Coulomb blockade charging steps of the box, 
strong enough  dissipation suppresses these fluctuations and leads to the reappearance of sharp 
charging steps. In the present paper we study 
this quantum phase transition in detail  using bosonization and numerical renormalization group methods in 
the limit of vanishing level spacing. 
\end{abstract}
\pacs{75.20.Hr, 71.27.+a, 72.15.Qm}
\maketitle

\section{Introduction}
Coulomb blockade is one of the most studied and most basic correlation 
effects that occurs in mesoscopic devices:\cite{Coulomb_blockade} 
This phenomenon appears in small structures (quantum dots or metallic islands)
weakly connected to the rest of the world. The capacitance of these  
small devices  can be very small and therefore the charging energy $E_C$, 
corresponding to the energy cost of putting an extra electron on the device,  can 
become large. If this charging energy is larger than the measurement 
temperature then correlation effects can become important, thermal fluctuations 
of the charge of the island become small, and transport through the 
device is typically suppressed as we lower the temperature.

In addition to the above-mentioned thermal fluctuations, 
Coulomb blockade can also be lifted by {\em quantum fluctuations}. There is two 
important known mechanisms that lead to quantum fluctuations: (a) Depending on the applied gate 
voltages, two charging states of the quantum dot can be degenerate, and electrons  
can thus gain kinetic energy by hybridizing with states in the leads, accompanied by strong quantum 
fluctuations of the charge.\cite{Matveev,Schoen} (b) Quantum fluctuations can play an important role 
at temperatures below the single particle level spacing of the dot too, if the dot contains 
an odd number of electrons or if the ground state of the isolated dot is degenerate.
In this case the system gets rid of the residual entropy associated with the spin 
degeneracy  of the isolated dot through the Kondo effect,\cite{Hewson}
which consists of binding conduction electrons in the attached metallic leads antiferromagnetically 
to the spin of the dot.\cite{Kondo_dot,David}

\begin{figure}
\epsfxsize8.0cm    
\epsfbox{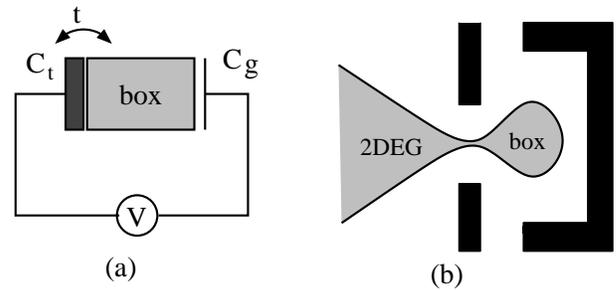}
\caption{
\label{fig:SEB} Sketch of the single electron box. Fig.~b. shows the top view of the 
 regime where electrons move in the the two-dimensional electron gas.  
Black areas indicate various  electrodes necessary to shape the electron gas. 
}
\end{figure}

The simplest device where charge fluctuations play a dominant role is the single electron box
(SEB) schematically shown in Fig.~\ref{fig:SEB}. In this device one couples an isolated quantum box
capacitively to a gate electrode and to an external lead through a tunnel junction. 
The isolated box can be very well described by the following simple Hamiltonian\cite{Aleiner_review}
\begin{equation}
H_{\rm box} = \sum_{n,\sigma} \epsilon_n d^\dagger_{n , \sigma} d_{n, \sigma} 
+ {e^2\over 2C}(\hat N -n_g)^2\;,
\label{eq:charging}
\end{equation} 
where $d^\dagger_{n, \sigma}$ creates an electron on the dot with spin $\sigma$ 
in a single particle state of energy $\epsilon_n$. The second term accounts for the 
electron-electron interaction with $C$ the total capacitance of the box, and
$\hat N = \sum_{n,\sigma} :d^\dagger_{n\sigma}d_{n,\sigma}:$ the number of extra electrons on 
the island.  In Eq.~(\ref{eq:charging})  $n_g = V_g C_g/e$ stands the dimensionless gate voltage 
 expressed in terms of the gate voltage $V_g$, the gate capacitance, $C_g$, and the
 electron charge $e$.  
Clearly, the number of electrons on an isolated  box jumps by one 
at approximately half-integer values of $n_g$ as we increase  $n_g$.\cite{note1}

In reality, however, the box is {\em not fully isolated} from the rest of the world, and 
in the vicinity of these 
degeneracy points quantum fluctuations to the leads need to be considered. 
In the following we shall restrict our considerations  to the particularly interesting regime 
where the temperature 
is much larger than the typical single particle level spacing $\Delta$ on the box, but is also 
much smaller than the charging energy $E_C = {e^2/2C}$, 
\be 
\Delta \ll T \ll E_C\;.
\ee
Furthermore, we shall focus our attention to the vicinity of the above  degeneracy points, 
$n_g \approx\;$half-integer.
As shown by Matveev,\cite{Matveev}  
in this regime quantum fluctuations of the charge on the dot 
can be treated by mapping the  Hamiltonian to that  of the anisotropic 
two-channel Kondo  problem. The finite jumps in $\langle \hat N\rangle$  are 
rounded by these quantum fluctuations, and 
all what is left from them  in the $\Delta\to 0$ limit is a {\em logarithmic divergence in the slope} 
of  $\langle \hat N\rangle(n_g)$ at the degeneracy points $n_g \approx\;$half-integer 
(see Fig.~\ref{fig:steps}.a). 

\bfig
\epsfxsize8cm
\epsfbox{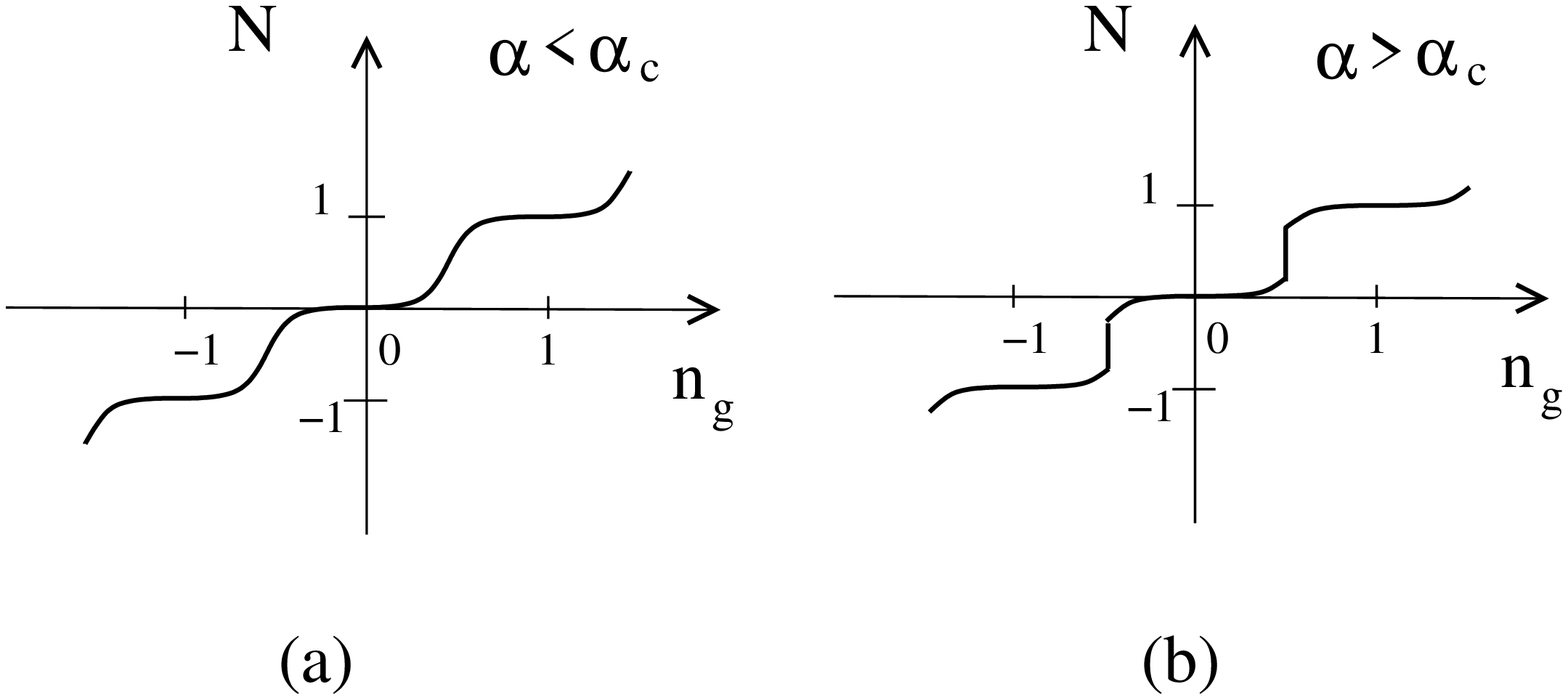}
\caption{\label{fig:steps}
The number of electrons on a single electron box as a function of the dimensionless
gate voltage. (a) The sudden jumps of an isolated box become smeared out due to quantum fluctuation
as soon as we couple the box to a lead. (b) Large enough dissipation restores the discrete jumps.}
\efig

However, as shown recently by Le Hur,  this picture is not complete:\cite{LeHur}  
 In  Eq.~(\ref{eq:charging}) one typically assumes that the gate voltage is constant. 
In reality, however, this gate voltage 
represents 
electromagnetic degrees of freedom. The fluctuations of these electromagnetic degrees of freedom 
 must be considered and typically result in Ohmic dissipation, characterized by a dimensionless dissipation 
strength $\alpha$.\cite{Leggett} The specific value of this dissipation 
depends on the resistance $R_g$ of the gate electrode and the gate capacitance, 
and is proportional to 
\be 
\alpha \approx {C_g^2\over 4 C^2} {R_g\over R_Q}\;,
\label{eq:physical_dissip}
\ee
with $R_Q = {h\over 2 e^2}$ the quantum resistance unit.
As shown by Le Hur, this  additional dissipative coupling leads to a quantum 
phase transition: Below a critical dissipation 
strength, $\alpha_c$, dissipation is ultimately irrelevant, and the scenario of Matveev 
holds with somewhat modified  parameters. If, however, $\alpha$ is large enough then dissipation 
dominates,  charge fluctuations are suppressed, and $\langle \hat N\rangle (n_g)$ displays 
{\em finite jumps} at half-integer values of $n_g$ at $T=0$ temperature, 
as shown in  Fig.~\ref{fig:steps}.b.\cite{LeHur} 
This critical value of dissipation depends on the dimensionless conductance $g$ of the  
tunnel junction, $\alpha_c = \alpha_c(g)$.

We have to remark here that the  above considerations hold only in the  limit of 
vanishing level spacing, $\Delta \to 0$. 
In reality, level spacing on the dot is finite, and for semiconducting devices 
the ratio $E_C/\Delta$ cannot be much larger than about $ 100$.\cite{2CKexp,Wilhelm}  
Thus  the level spacing $\Delta$ ultimately provides a finite cut-off similar to 
a finite temperature,  and the slope of  $\langle \hat N\rangle (n_g)$ is expected 
to remain finite  even at $T=0$ temperature. Although it is not clear what happens below the level spacing, 
in a real system the phase transition above is thus probably  reduced to a cross-over
(see our conclusions too).

In the present paper we shall investigate this phase transition employing 
non-perturbative methods such as Abelian bosonization\cite{JanSchoeller} and numerical 
renormalization  group to a somewhat generalized version of the 
quantum box model.\cite{wilson75,costi94} These methods enable us to obtain the full 
phase diagram of the quantum box. 

Using  the bosonization approach we are able to construct the non-perturbative 
scaling equations that describe the above quantum phase transition. 
At the degeneracy point, we find a Kosterlitz-Thouless    
quantum phase transition,  in agreement with the weak coupling result of 
Le Hur.\cite{LeHur} We also determine the scaling of the 
Kondo temperature $T_K$ associated with  the charge fluctuations non-perturbatively.
Apart from an overall  prefactor,  this scale vanishes as 
\be 
T_K(\alpha)  \approx E_C \;
{\rm exp}\Bigl\{-{\pi \over \sqrt{{2\over \pi} \sqrt{g} (1-\alpha_c)^{1/2}} 
\sqrt{\alpha_c - \alpha}}\Bigr\}\;,
\ee
where $g= {h\over 2e^2} G$ is the dimensionless conductance of the junction.
Thus $T_K$  approaches zero very fast but continuously as
one approaches the critical value $\alpha_c$ from below.

The phase transition also manifests itself in the $T=0$ temperature 
linear conductance of the tunnel junction, which displays a first order phase 
transition:  
\be
g(T=0,\alpha) = {1\over 2} \theta(\alpha_c-\alpha)\;.
\ee

We also study the above phase transition using the powerful method 
of numerical renormalization group (NRG). Since the calculation for fermions 
with spin is computationally extremely demanding, we will have to restrict ourself
to the case of spinless  fermions. 
The phase transition within this method simply appears as 
a change in the finite size spectrum.  The NRG  method also enables us 
to determine the  shape of the step  in $\langle \hat N\rangle (n_g)$
and $T_K$ non-perturbatively, and compute various spectral functions. 
Our results agree very well with the scenario of Ref.~\onlinecite{LeHur}
and the bosonization results.

The paper is structured as follows: In Section~\ref{sec:Hamilt} we introduce 
our basic notations and show how to map the problem to a Kondo model.
Section~\ref{sec:bosnize} is devoted to the construction of the non-perturbative
scaling equations and their analysis. The numerical renormalization 
group results are presented in Section~\ref{sec:NRG}. Finally, we summarize 
our results and discuss future perspectives in 
Section~\ref{sec:conclusions}

\section{Hamiltonian}
\label{sec:Hamilt}

We shall 
assume that coupling between the lead and the dot is small, so that charge fluctuations 
can be described within the tunneling approximation 
\be 
H_{\rm tun} = t \sum_{n, \epsilon,\sigma} (d^\dagger_{n,\sigma}c_{\epsilon,\sigma} + {\rm h.c.})\;,
\label{eq:H_tun}
\ee
where the operators $ c_{\epsilon,\sigma}^\dagger$ create conduction electrons in the 
lead with energy $\epsilon$ and spin $\sigma$,
\be 
H_{\rm lead} = \sum_{ \epsilon,\sigma}\epsilon \; c_{\epsilon,\sigma}^\dagger c_{\epsilon,\sigma} \;.
\ee
In Eq.~(\ref{eq:H_tun})
we assumed that there is only a single tunneling mode that is coupled to the 
dot electrons so that the tunneling matrix elements can be replaced by a simple number. 
The situation is much more complex in case of several tunneling modes
and shall not be discussed here (see Ref.~\onlinecite{Wilhelm}).

Since all half-integer values of $n_g$ are essentially equivalent, 
to  obtain Matveev's  mapping, we shall  focus to the vicinity of the point 
$n_g = 1/2$. 
The important quantum fluctuations in this regime are dominated by the two 
charging states $|N=0\rangle$ and $|N=1\rangle$ of the dot.  Other charging states of the dot 
are separated at least by an energy $\sim E_C$, and in the small temperature/tunneling regime
their effect is only to slightly renormalize the values of $t$ and $E_C$.  We can 
thus restrict our considerations to the charging states $N=0$ and $N=1$, and
keep track of  charge fluctuations of the box by introducing an {\em orbital pseudospin} $T$, 
the states $T_z=(1-\hat N)/2 = \pm1/2$ corresponding to states with $N=0,1$. Note that we still have to 
keep track of internal electron-hole excitations of the box, since our energy scale is larger than 
the level spacing.

To rewrite the tunneling Hamiltonian in a more suggestive way 
we introduce the new fields normalized by the density of states in the box
and in the lead, $\varrho_{\rm box}$ and  $\varrho_{\rm lead}$, respectively
\be
D_{\sigma} \equiv {1\over \sqrt{\varrho_{\rm box} }} \sum_{n} d_{n,\sigma}\;,
\phantom{nn}
C_{\sigma} \equiv {1\over \sqrt{\varrho_{\rm lead} }} \sum_{\epsilon} d_{\epsilon,\sigma}\;,
\label{eq:fields}
\ee 
and organize them into a four component spinor
\be
\psi_{\tau,\sigma} \equiv \pmatrix{C_\sigma  \cr D_\sigma \cr}\;. 
\ee
The normalization in Eq.~(\ref{eq:fields}) has been chosen is such a way that the imaginary time 
propagator of the field  $\psi$ satisfies
$
\langle {\rm T}_\tau \psi_{\tau\sigma}(\tau) \psi_{\tau'\sigma'}^\dagger(0)\rangle = 
{\delta_{\sigma,\sigma'} \delta_{\tau,\tau'}/  \tau}\;.
$
In this language  we can rewrite the tunneling part of Hamiltonian as
\be 
H_{\rm perp} =  {j_\perp \over 2}   ( T^+ \psi^\dagger_{\sigma} \tau^- \psi_{\sigma} + {\rm h.c.})\;,
\label{eq:H_tun2}
\ee
where the operator $\tau^\pm$ just flips the orbital spin $\tau$ of the field $\psi_{\tau\sigma}$, and 
the $j_\perp$  is a dimensionless coupling proportional to the tunneling, 
$j_\perp = 2 t \sqrt{\varrho_{\rm box} \varrho_{\rm lead}}$. Thus $j_\perp^2$ is directly related to 
the dimensionless conductance $g$ of the tunnel junction
\be
g = {G\over G_Q} = {\pi^2\over 4} j_\perp^2\;,
\ee
with $G_Q = 2e^2/h$, the quantum conductance unit.
Clearly, Eq.~(\ref{eq:H_tun2}) is just the 
Hamiltonian of an anisotropic two-channel Kondo model,\cite{Matveev,Cox}  the orbital spins $T$ and $\tau$ 
playing the role of the spins of the original two-channel Kondo model, and the electron spin
$\sigma$ providing the silent channel index. The presence of this additional channel index 
makes the physics of the two-channel Kondo model entirely different from that of the 
single channel Kondo problem, and leads to  non-Fermi liquid properties, such as the 
logarithmically divergent capacitance in our case. We emphasize again, that the above
 mapping holds only in the  regime $\Delta< T,\omega,... < E_C$, where the level spacing of 
the box can be neglected.

Deviations from the point $n_g=1/2$ appear in this language as a local field acting on the 
orbital spin of the box
\be
{e^2\over 2C}(\hat N -n_g)^2\; \to \; - B T^z\;,
\label{eq:deviation}
\ee 
where the effective magnetic field is simply given by $B =  E_C (1 - 2 n_g)$.
It is clear from this expression that fluctuations of the gate voltage, $\delta V_g$ 
will result in a  dissipative coupling of the following form 
\be
H_{\rm diss} = \lambda \;  T^z\; \varphi ,
\ee
where for an Ohmic heat bath the bosonic field $\varphi$ decays as
\be
\langle {\rm T}_\tau \varphi(\tau) \varphi(0)\rangle = {1\over \tau^2}\;.
\label{eq:fluc}
\ee
The dimensionless coupling $\lambda$ above is related to the usual coupling 
$\alpha$ as $\alpha = \lambda^2/2$. The specific value of $\lambda$ depends on 
the properties of gate electrodes and is approximately given by Eq.~(\ref{eq:physical_dissip}).
Thus for very resistive 
gate electrodes, $\alpha$ can be large, while if the gate electrodes are 
good conductors, then $\alpha$ is small.

Eqs.~(\ref{eq:H_tun2}), (\ref{eq:deviation}), and (\ref{eq:fluc}) constitute the effective Hamiltonian 
that describes the physics of the dot in the vicinity of the degeneracy point 
in the presence of dissipation.
In the rest of the paper we shall, however,  include an additional term in the Hamiltonian
of the form
\be 
H_{z} =  {j_z \over 2} \; \sum_\sigma  \;T^z \; \psi^\dagger_{\sigma} \tau^z \psi_{\sigma} \;
\label{eq:H_z}
\ee
too.
This term describes charging state dependent scattering of the conduction electrons.
While this term is typically small, it is generated under the renormalization 
group procedure. Furthermore, in other physical systems such as
a Kondo spin coupled to fluctuations of quantum critical spin
degrees of freedom, {\em e.g.},\cite{Demler}
this term is present in the bare Hamiltonian as well.

\section{Bosonization}
\label{sec:bosnize}

The non-perturbative method of Abelian bosonization can be very efficiently used to 
construct the phase diagram of our model, and allows us to treat 
both the dissipative coupling $\lambda$ and the coupling $j_z$ non-perturbatively. 
For the sake of simplicity, we shall first  
discuss the somewhat simpler case  of spinless Fermions and then restore spin indices. 
Throughout this Section we shall use the bosonization scheme
 of Ref.~\onlinecite{JanSchoeller}.

\subsection{Bosonization for spinless Fermions}

For spinless fermions the field $\psi$ has only two orbital indices, 
$\tau = \pm$. Correspondingly, we have to introduce two left-moving 
bosonic fields $\Phi_\pm$ to represent $\psi_\pm$,\cite{JanSchoeller}
\be 
\psi_\pm  = F_\pm {1\over \sqrt{a}} e^{-i \Phi_\pm}\;,
\ee
where $F_\pm$ denotes the Klein factor, and $a$ is a short distance 
cut-off $\sim v_F /E_C$.  In terms of these bosonic fields the interaction 
part of the Hamiltonian reads
\bea 
H_z &  = &  {{\tilde j}_z  \over \sqrt{2}} \;T^z \; \partial_x \Phi_T\;,
\label{eq:bos1}
\\
H_{\rm perp} & = & 
 j_\perp {1\over 2 a}   ( T^+ e^{-i \sqrt{2}\; \Phi_T }  + {\rm h.c.})\;,
\eea
where for the sake of simplicity we suppressed Klein factors and introduced the orbital field, 
$\Phi_T = (\Phi_+ - \Phi_-)/\sqrt{2}$. Note that the charge field 
$\Phi_C = (\Phi_+ + \Phi_-)/\sqrt{2}$ completely decouples from 
the orbital spin of the box. 
Particular care is needed  while
 deriving Eq.~(\ref{eq:bos1}): Introducing the 
bosonization cut-off scheme  renormalizes the coupling constants, 
$j_z\to {\tilde j}_z$. 
\be 
{\tilde j}_z = {4\delta_z\over \pi} = {4\over \pi} {\rm atan}
\left({\pi j_z\over 4}\right)\;, 
\ee
with  $\delta_z$ the phase shift generated by $j_z$. Representing 
the field $\varphi$ as the derivative of another bosonic field
$\Phi$ we can finally rewrite the dissipative term as 
\be 
H_{\rm diss} = \lambda T^z {\partial_x \Phi}\;. 
\ee
In this bosonized form the non-interacting part of the Hamiltonian becomes
\be 
H_0= \int {dx\over 4\pi} \Bigl[\sum_{\mu = T,C} :(\partial_x \Phi_\mu)^2:
+  :(\partial_x \Phi)^2: \Bigr]\;,
\ee
where the first two terms represent electron-hole excitations on the 
dot and in the lead, and the last term gives the energy of
 excitations responsible for dissipation. 


To obtain the phase diagram of our model we carried out a non-perturbative scaling 
analysis. As a first step we eliminated the terms proportional to 
$\lambda$ and $j_z$ by applying a unitary transformation
\be 
U = {\rm exp} \Bigl\{ i \;T_z \;\bigl(
 \lambda \Phi(0) + {4\delta_z\over \pi\sqrt{2}} \Phi_T(0)\bigr)\Bigr\}\;.
\ee
This unitary  transformation brings $H_{\rm perp}$ to the following form
\be 
H_{\rm perp} = {j_\perp\over 2a} \Bigl[ T^+ 
e^{ i 
 \lambda \Phi(0) + i \bigl( {4\delta_z\over \pi\sqrt{2}} - \sqrt{2} \bigr) 
\Phi_T(0)} + {\rm h.c.} \Bigr]\;.
\ee

The scaling dimension of $j_\perp$ can be read out of this equation 
and one obtains 
\be 
{dj_\perp\over dl} = \Bigl[ 4{\delta_z\over \pi} 
- 4\bigl({\delta_z\over \pi}\bigr)^2 - \alpha \Bigr]
j_\perp + {\cal O}(j_\perp^3)\; ,
\ee
where $\alpha = \lambda^2/2$ is the dissipation strength, and $l=\ln\; a$ is the scaling 
variable. The first term in this equation is just the famous non-perturbative equation 
of Yuval and Anderson.\cite{YuvalAnderson}  The second term in this scaling equation 
just means that dissipative coupling to charge fluctuations 
on the gate tends to suppress quantum fluctuations of the charge
of the dot. It is this term that drives the phase transition. 

The scaling equations of $\alpha$ and $\delta_z$ can be simply obtained 
by taking the operator product expansion of the term $j_\perp$ with itself. 
In this way we obtain:
\bea 
&&{d\over dl} \Bigl( {4\delta_z\over \pi} \Bigr) = 
\Bigl(1 
- {2\delta_z\over \pi}\Bigr) \; j^2_\perp\;,
\\
&&{d\alpha\over dl} = -\alpha j_\perp^2\;,
\eea
where terms of order $j_\perp^3$ have been dropped. 
The first of these equations has been derived  by Yuval and Anderson,\cite{YuvalAnderson} 
and shows that $j_\perp$ tends to boost up $\delta_z$ and generate a Kondo effect. 
The  second equation shows that quantum fluctuations {\em suppress}
dissipation effects. There is thus a competition between $j_\perp$ and $\alpha$, 
which mutually tend to suppress each-other. 

Depending on the model parameters, 
we have to distinguish two regimes: In the small dissipation regime  dissipation is irrelevant, 
$\alpha$ scales to zero, and charge fluctuations generated by $j_\perp\sim t$
start to dominate at an energy scale $T_K$, the so-called Kondo energy. 
This Kondo fixed point corresponds to  $\delta_z = \pi/2$ and 
$j_\perp\to {\cal O}(1)$.\cite{YuvalAnderson} 
It is easy to show that the dissipative coupling is indeed irrelevant 
at this Kondo fixed point: This follows from the observation that 
the Kondo fixed point is described by a Fermi liquid theory.\cite{Hewson} 
As a consequence,  the orbital susceptibility is finite at zero temperature, 
$\chi_{T}\sim 1/T_K$, and correspondingly the imaginary time correlation 
function of $T_z$ decays as $\langle {\rm T}_\tau  
T_z(\tau) T_z(0)  \rangle \sim 1/\tau^2$. Therefore
the correlation function $\langle {\rm T}_\tau  
T_z(\tau) \varphi(\tau) T_z(0) \varphi(0) \rangle$ decays as  $1/\tau^4$,
implying that the scaling dimension of the dissipation is $-1$ at 
the Kondo fixed point, and is therefore irrelevant.\cite{note_on_dimensions}  
The capacitance of the dot for these small values of $\alpha$ 
is simply proportional to the orbital 
susceptibility, $C\sim \chi_T\sim 1/T_K$.

If, on the other hand, the bare value of $\alpha$ is larger
than a critical value, $\alpha_c(j_\perp,\delta_z)$, then 
quantum fluctuations  are suppressed, and $j_\perp$ scales to 0
implying that charge fluctuations are suppressed. 
This phase can be called {\em localized} in the sense that 
if we prepare the box in the state $\hat N = 0$ at the degeneracy point,
then the expectation value of $\hat N$ will never approach its equilibrium value
$1/2$, 
$$
\lim_{t\to\infty} \langle \hat N (t) \rangle < 1/2\phantom{nnn}
(\alpha>\alpha_c).
$$ 
This behavior is analogous to that of the dissipative 
two-state system.\cite{Leggett} 

It is easy to determine this critical value of $\alpha$ in the 
$j_\perp\to 0 $ limit from our scaling equations:
\be 
\alpha_c(j_\perp\to0) = 4 {\delta_z\over \pi} \Bigl(
1- {\delta_z\over \pi}\Bigr)\;.
\label{eq:alpha_c_jperp0}
\ee
To obtain the phase diagram in the vicinity of this line.
we rewrite the scaling equations as
\bea 
&&{dj_\perp \over dl} \approx j_\perp \;r\;,
\\
&&{d r\over dl} = j_\perp^2\;,
\eea
where we introduced the variable
\be 
r =  4 {\delta_z\over \pi} \Bigl(
1- {\delta_z\over \pi}\Bigr) - \alpha\;,
\ee
measuring the distance from the line Eq.~(\ref{eq:alpha_c_jperp0}).
These equations clearly describe a Kosterlitz-Thouless phase transition.\cite{KosterlitzThouless} 
Solving these equations we  find that for small 
values of $j_\perp$
\be
\alpha_c \approx |j_\perp| + 4 {\delta_z\over \pi} \Bigl(
1- {\delta_z\over \pi}\Bigr)\;,
\label{eq:alpha_c_spinless}
\ee
and the corresponding Kondo scale vanishes as one approaches 
$\alpha$ as
\be 
T_K \approx E_C \;
{\rm exp}\Bigl\{-{\pi \over \sqrt{2|j_\perp|} \sqrt{\alpha_c - \alpha}}\Bigr\}\;.
\label{eq:TK_nospin}
\ee

It is important to remark at this point that the above scaling equations describe the 
nature of the phase transition 
even for {\em large} values of $j_\perp$. The reason is that in the vicinity of the 
quantum phase transition {\em all scaling trajectories approach the 
critical line} 
defined by $j_\perp=0$ and $\alpha =  4 {\delta_z\over \pi} [
1- {2\delta_z\over \pi}]$. Thus the initial part of the scaling just 
renormalizes the effective value of the various couplings
that enter Eq.~(\ref{eq:T_K^spin}), but the overall picture does not change.

\subsection{Fermions with spin}

The considerations of the previous subsection can be easily generalized 
to the case of Fermions with spin. The scaling equation 
of $j_\perp$ then becomes
\be 
{dj_\perp\over dl} = \Bigl[ 4{\delta_z\over \pi} 
- 8\bigl({\delta_z\over \pi}\bigr)^2 - \alpha \Bigr]
j_\perp\;,
\label{eq:j_perp_scaling_spinful}
\ee
while the other two scaling equations read
\bea 
&&{d\over dl} \Bigl( {4\delta_z\over \pi} \Bigr) = 
\Bigl(1 
- {4\delta_z\over \pi}\Bigr) \; j^2_\perp\;,
\\
&&{d\alpha\over dl} = -\alpha j_\perp^2\;.
\eea
In the absence of dissipation, these scaling equations have been 
derived first by Vlad\'ar, Zim\'anyi and Zawadowski.\cite{VladZimZaw} 
Similar to the spinless case, two regimes can be distinguished:\cite{LeHur} 
For small dissipations $\alpha$ is irrelevant, and 
at the degeneracy point of the dot the couplings 
$j_\perp$ and $\delta_z$ flow to the two-channel Kondo fixed point, 
$\delta_z = \pi/4$, $j_\perp \to {\cal O}(1)$.
The physics is, however, not that of a Fermi liquid due to the presence 
of electron spin: 
the orbital susceptibility ({\em i.e.} the capacitance) of the dot 
diverges logarithmically below the Kondo temperature, 
$\chi_{T}\sim {\rm ln}(T_K/T)/T_K$.\cite{Cox,Matveev}
This logarithmic divergence indicates that the deviation 
from the degeneracy point $n_g=1/2$ represent a relevant 
perturbation of scaling dimension $1/2$, and as a consequence, 
the slope of $\langle \hat N\rangle(n_g)$
diverges logarithmically at $T=0$ temperature as one approaches 
$n_g=1/2$ (if we neglect the finite level spacing on the dot).

It is easy to check that dissipation is  irrelevant at this two-channel 
Kondo fixed point too: The logarithmic divergence of the orbital 
susceptibility corresponds to an asymptotic decay 
$\langle {\rm T}_\tau  T_z(\tau) T_z(0)  \rangle \sim 1/\tau$.
Consequently, the correlation function $\langle {\rm T}_\tau  
T_z(\tau) \varphi(\tau) T_z(0) \varphi(0) \rangle$ decays as  $1/\tau^3$
at the two-channel Kondo fixed point, 
implying that the scaling dimension of the dissipation is $-1/2$ at 
the Kondo fixed point, and is therefore irrelevant.\cite{note_on_dimensions}  

If, however, $\alpha$ is large enough then again, the dissipation 
drives $j_\perp$ to zero, and we recover the {\em localized} phase 
with {\em jumps} in the function $\langle \hat N\rangle(n_g)$.
Similar to the spinless case, we can determine the boundary between these 
two phases by introducing the  variable 
\be 
r = 4 {\delta_z\over \pi} \Bigl(
1- {2\delta_z\over \pi}\Bigr)-\alpha\;,
\ee
and rewriting the scaling equations in terms of it. Keeping only the lowest 
order terms in $r$ and $j_\perp$ we obtain again a set of two coupled 
differential equations describing a Kosterlitz-Thouless phase transition:
\bea 
&&{dj_\perp \over dl} \approx j_\perp \;r\;,
\\
&&{d r\over dl} = (1-\alpha) j_\perp^2\;.
\eea
For small values of $j_\perp$ the phase boundary is thus given by
\be
\alpha_c^{\rm spin} \approx j_\perp + 4 {\delta_z\over \pi} \Bigl(
1- {2\delta_z\over \pi}\Bigr)\;,
\ee
and the  Kondo scale vanishes as one approaches 
$\alpha_c$ from below as
\be 
T_K  \approx E_C \;
{\rm exp}\Bigl\{-{\pi \over \sqrt{2|j_\perp| (1-\alpha_c^{\rm spin})^{1/2}} 
\sqrt{\alpha_c^{\rm spin} - \alpha}}\Bigr\}\;.
\label{eq:T_K^spin}
\ee

\section{Numerical renormalization group calculation}
\label{sec:NRG}

In order to obtain results for different physical quantities we
performed 
Wilson's numerical renormalization group method (NRG) on the model.
\cite{wilson75,costi94,bulla97,hofstetterDMNRG} 
This method is especially suitable for quantum impurity 
problems and can be used to determine dynamical as well as thermodynamic 
properties with high accuracy. 
However,  performing such a calculation on the Bose-Fermi Kondo model, a
conceptual problem arises: Originally, NRG was designed to treat
{\em fermionic} problems, where the method is based on a logarithmic
discretization of the conduction band and then a mapping 
of it onto a semi-infinite chain. In the present
model we need to treat both fermionic and bosonic baths at the same time. 
Even though the NRG method has recently been extended to bosonic
systems,\cite{bulla03,bulla04} the bosonic calculation 
needs considerable computational effort. We therefore decided to treat 
this problem in a different way. 

Fortunately, in the present model the bosonic field $\varphi$ is Ohmic. 
This implies that we can represent this field by coupling $T_z$
to the total density of $F$ fermionic fields 
in the following way:
\be
H_{\rm diss} = {g\over 2} T^z \sum_{i=1}^F \Psi^\dagger_i \Psi_i\;.
\ee
Here the fields $ \Psi_i$ denote some fermion fields
normalized as $\langle {\rm T}_\tau \Psi_i(\tau) \Psi_j^\dagger(0)\rangle = 
\delta_{i,j}/ \tau$.  To make this mapping complete, we have 
to find the relation between $q$ and the dissipation strength
$\alpha$. It is a relatively simple matter to show that these are
connected as
\be 
\alpha = 2 F \left({\delta_g\over\pi}\right)^2 \;, 
\label{eq:alpha(g)}
\ee
with $\delta_g = {\rm atan}( {\pi g/ 4})$ the phase shift associated with the 
potential scattering $g$. 

The advantage of this trick is that we can use just fermionic fields 
in our calculations. On the other hand, this trick also implies that 
all along the calculations, we carry along  $F-1$ bosonic fields
which are decoupled from $T_z$. For large values of 
$F$ we thus keep track of many redundant degrees of freedom.
Furthermore, we cannot reach any value of $\alpha$ in this 
way. In our calculations, {\em e.g.},  we used $F=2$ fermion fields allowing 
to map out the phase diagram for $0\le\alpha \le 1$.

It is clear from the previous subsection that to represent the
physical system considered in the present paper, one  needs
at least {\em three} bands of spinful fermions: one of them, $\Psi_\sigma$  would be used 
to  represent dissipative term in the Hamiltonian while 
two others are just the electronic degrees of freedom, $\psi_{\tau\sigma}$. 
This implies that one needs a three-channel NRG
code to attack the problem in its full glory. Unfortunately, 
we do not have too many symmetries simplifying our problem: 
Throughout the calculation we used  an SU(2) symmetry associated with 
particle-hole symmetry and a U(1) symmetry associated with the  
$z$-component of the orbital spin.  As a consequence, 
we were  able to study  only the case of spinless fermions with sufficient 
accuracy.  Fortunately, as shown in the previous section, most features 
of the spinful case already appear in the spinless version of the model. 
Furthermore, the spinless case is also of relevance: 
this is realized in a single electron box by
applying an external magnetic field that lifts the electron-spin
degeneracy and drives the system away from the unstable
two-channel Kondo fixed point. 

\begin{figure}
\epsfxsize6.0cm    
\epsfbox{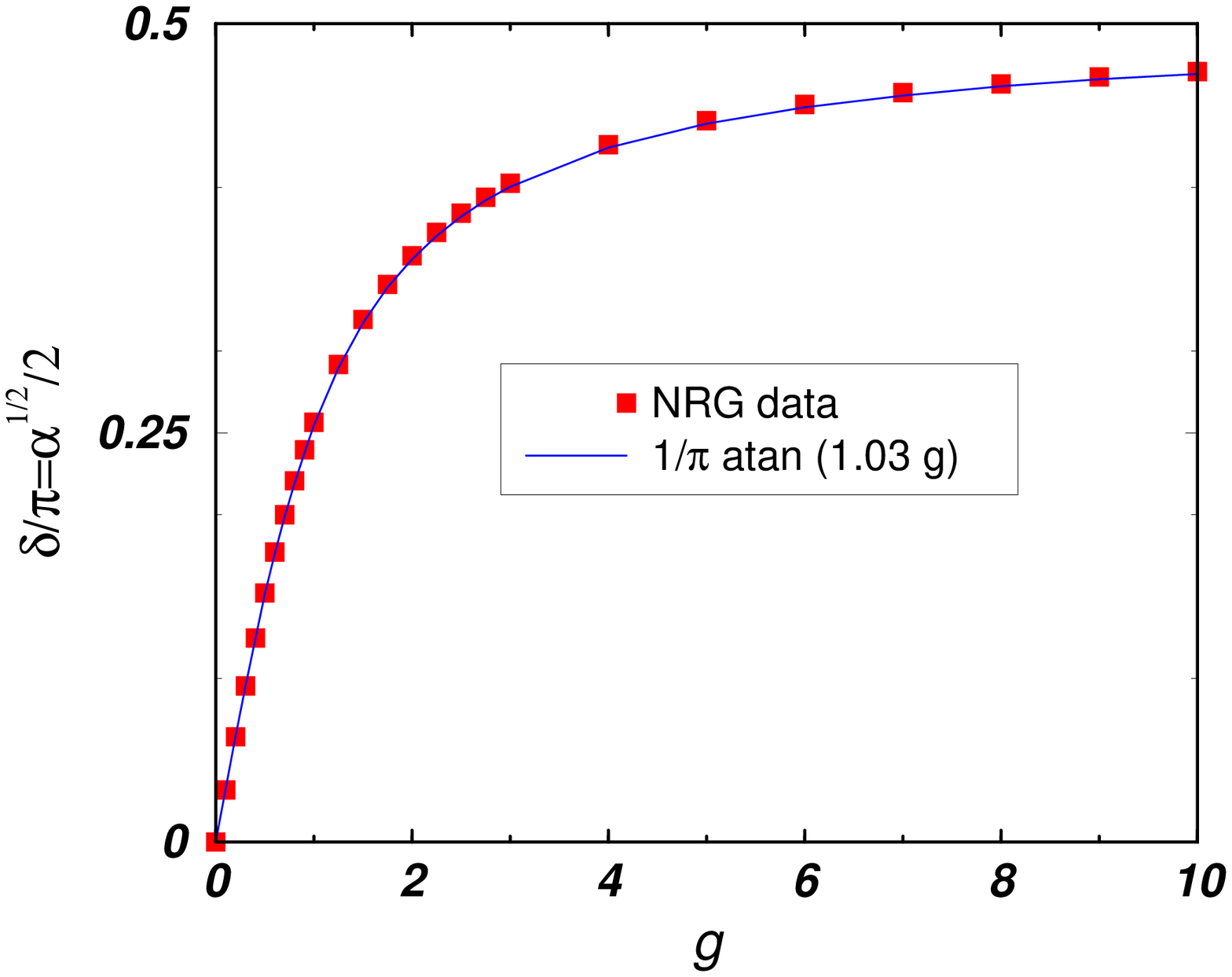}
\caption{
\label{fig:calib} 
(color online) Relation between the phase shift $\delta_g$ extracted from the finite size spectrum and the 
fermionic coupling $g$. The continuous line shows the fit with Eq.~(\ref{eq:delta_g}).
}
\end{figure}

The very first step is to calibrate the method, {\em i.e.},  to determine
the correspondence between the 
fermionic coupling $g$  and $\alpha$. In principle, these are related by
Eq.~(\ref{eq:alpha(g)}). However, the NRG discretization renormalizes the 
effective value of $g$.  To do that, we have performed
calculations for $j_\perp = j_z = 0$ in the presence of a term 
$\Delta_0 T^x$, {\em i.e.}, for the spin-boson problem, 
and extracted the phase shifts from the finite size spectra.
In that way we found that the phase shifts can be expressed 
 with very high accuracy as 
\begin{equation}
\delta_g ={\rm atan}(f(\Lambda)\;g)\;,
\label{eq:delta_g}
\end{equation}
where $\Lambda$ is the parameter of the logarithmic discretization
used in NRG and $f(\Lambda)$ is a 
factor
close to unity that must be determined numerically. 
In particular, for  $\Lambda=2$ used throughout this paper 
we find $f(\Lambda=2)=1.03$. 
The phase shift above can then be plugged into Eq.~(\ref{eq:alpha(g)})
to obtain the corresponding value of $\alpha$. This procedure is illustrated in Fig.\ref{fig:calib}. 
A similar expression 
gives the value of $\delta_z$ as a function of $j_z$.

\begin{figure}
\epsfxsize=6cm
\epsfbox{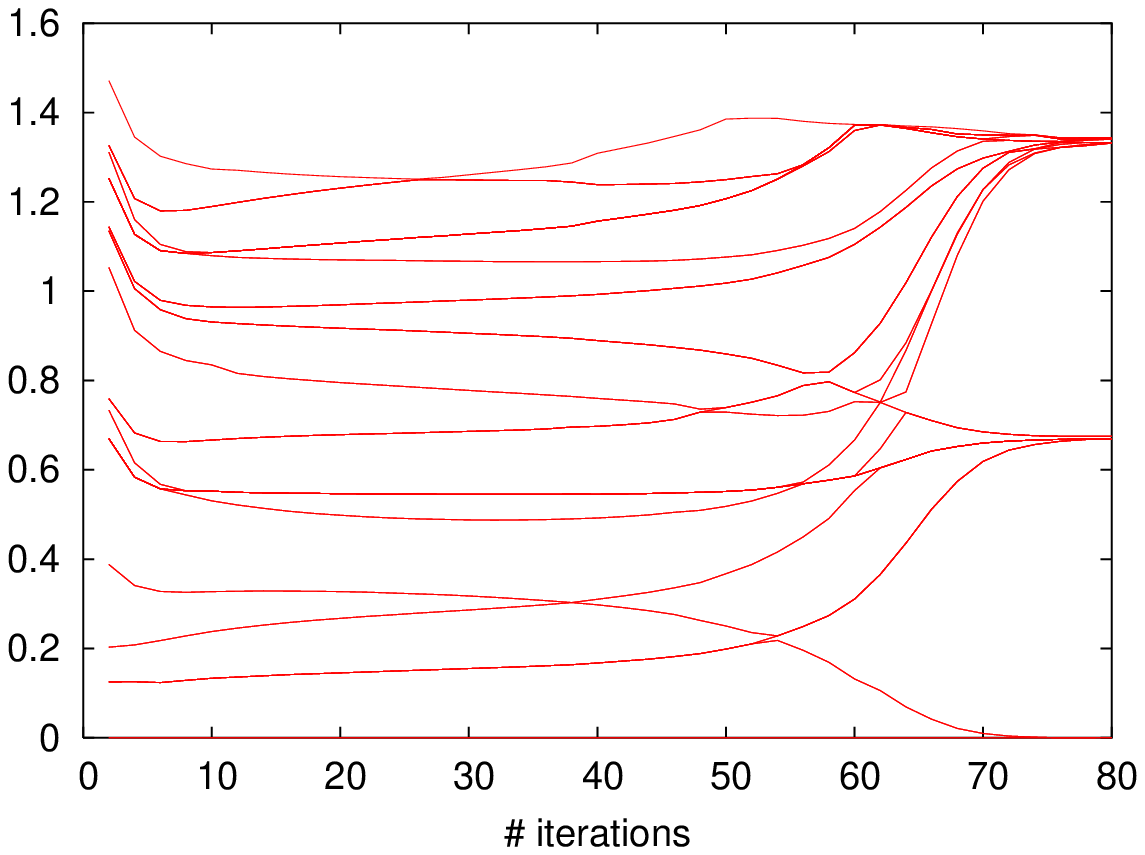}
\epsfxsize=6cm
\epsfbox{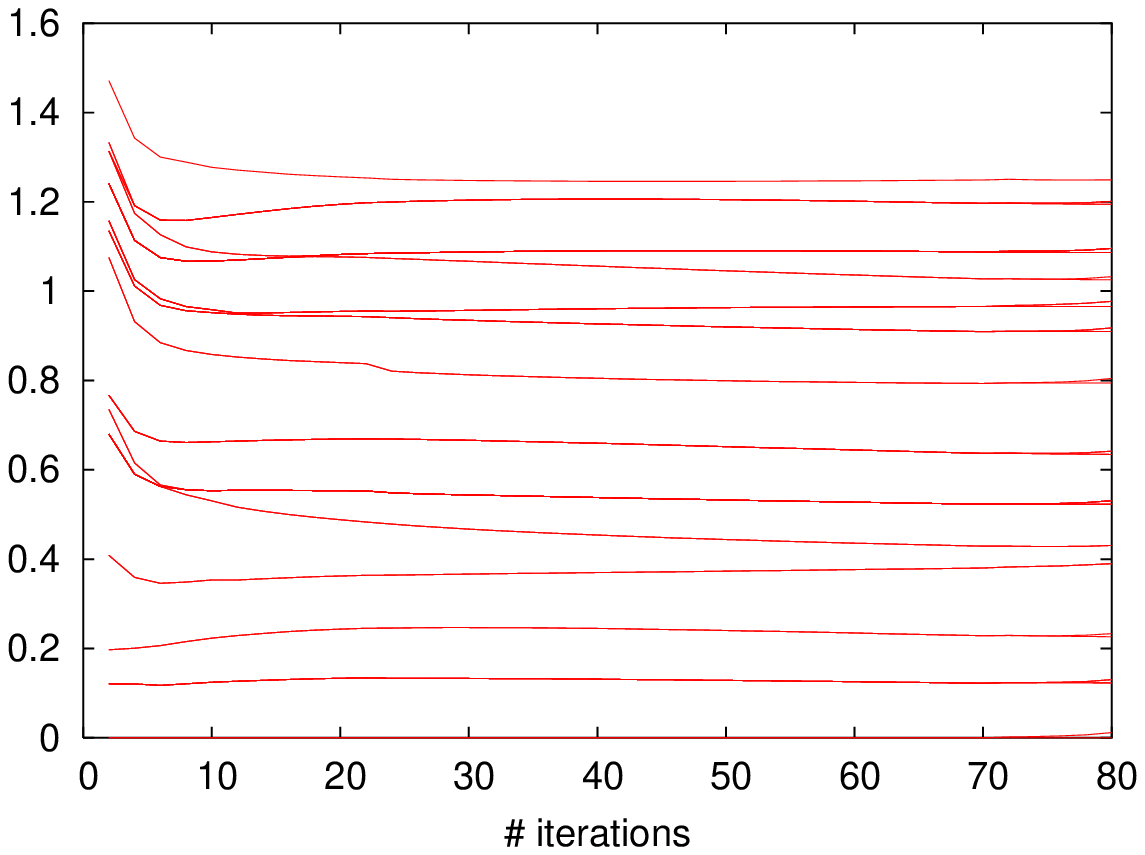}
\caption{(color online) Finite size spectra obtained by NRG for the two distinct
regimes. The upper part shows the spectrum for $\alpha=0.653$, $j_z=0.2$,
$j_\perp=0.7$, while the lower part stands for the same $j$s but for
$\alpha=0.667$. It is transparent that while on the left figure the
levels crosses over to the well-known Kondo spectrum, on the right
panel no such crossover occurs.} \label{fig:spectra}
\end{figure}

\begin{figure}
\epsfxsize=6cm
\epsfbox{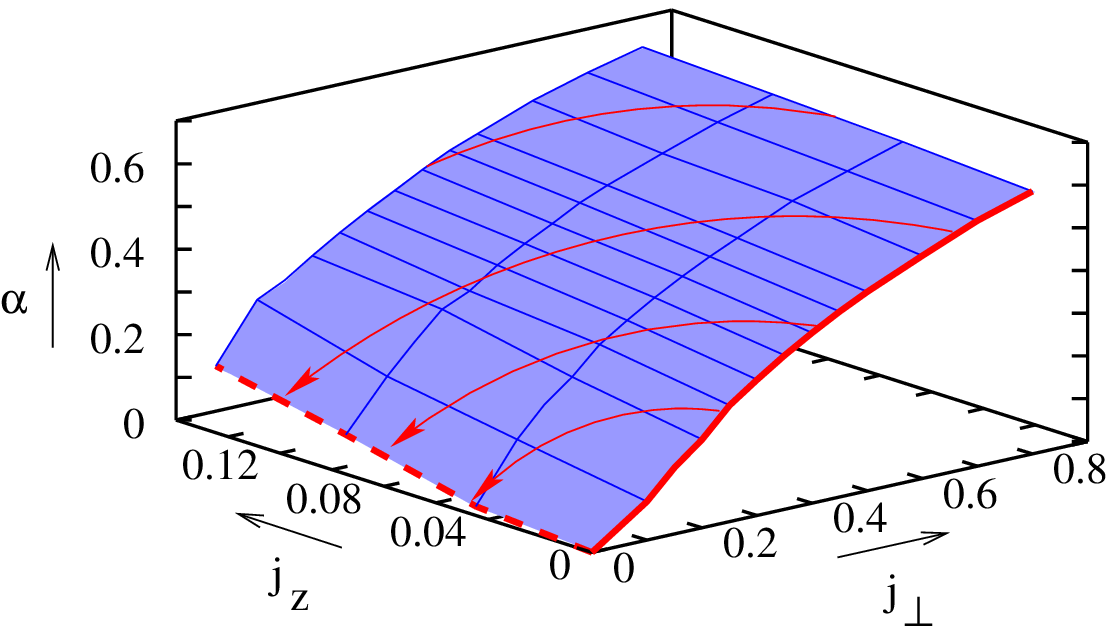}
\caption{(color online) The critical surface in the parameter space separating
the Kondo regime from the localized regime. The heavy continuous line indicates 
the phase transition line for the case relevant for the single electron box $j_z=0$.
The heavy dashed line indicates the {\em critical line}, $j_\perp = 0$. 
The arrows show the scaling trajectories for $\alpha=\alpha_c$.
All these critical scaling tranjectories end up at the critical line.}
\label{fig:critical_surface}
\end{figure}

The most basic result one can obtain by NRG is the finite size
spectrum, {\em i.e.},  the rescaled lowest lying energy levels plotted against
the iteration number. We show two typical spectra in
Fig.\ref{fig:spectra}. In the top panel of
Fig.\ref{fig:spectra} the spectrum crosses over to that of the usual Kondo
fixed point\cite{wilson75} with uniform level spacing
at around $N\sim60$.
In the bottom panel,
however, no such crossover occurs, at least not before the 80th
iteration indicating that the Kondo effect is suppressed by the
strong coupling to the bosonic bath which tends to suppress charge fluctuations 
and thus localizes the orbital spin of the dot. 
This robust difference in the finite size spectra allows us to extract
the critical value of the bosonic coupling $\alpha_c$ and map out the phase boundary 
for all values of $j_\perp$ and $\delta_z$. 
Numerically we say that we are in the localized (dissipation dominated) phase 
if the crossover to the Kondo spectrum does
not occur within $N=80$ iterations.
The critical surface corresponding to this phase transition is 
shown in Fig.(\ref{fig:critical_surface}). 

\begin{figure}
\epsfxsize=6cm
\epsfbox{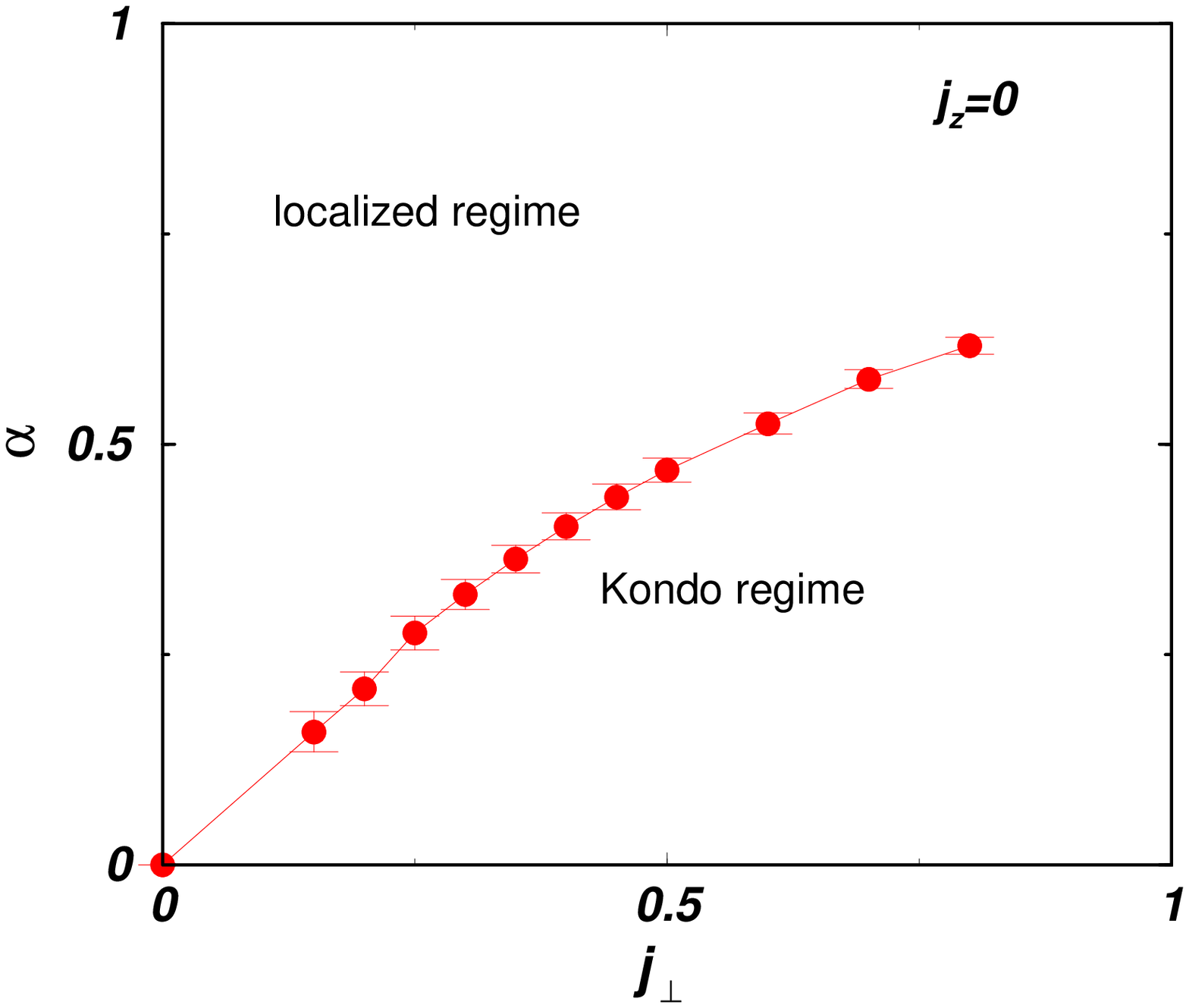}
\epsfxsize=6cm
\epsfbox{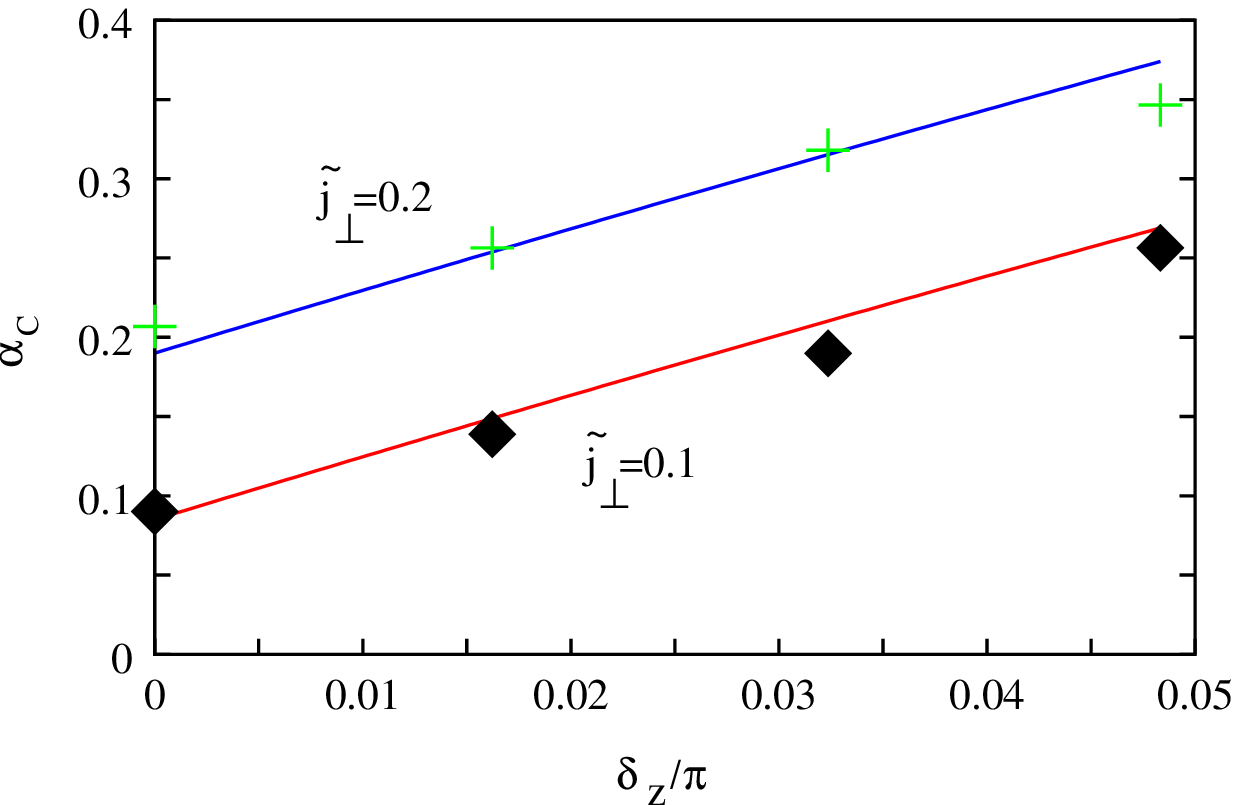}
\caption{Top: (color online) The physically relevant $j_z=0$ cut of the critical
surface. Bottom: Phase boundary as a function of $\delta_z/\pi$ for 
two values of $j_\perp$. These phase boundaries compare very well with 
the analytical expression Eq.~(\ref{eq:alpha_c_spinless}).}
\label{fig:critical}
\end{figure}

The case $j_z = 0$ is of particular interest, since this corresponds to the
charge fluctuations at the degeneracy point of the quantum dot. 
In agreement with Eq.~(\ref{eq:alpha_c_spinless}) $\alpha_c$ increases approximately linearly with
$j_\perp\sim t\sim \sqrt{g}$ see Fig.~\ref{fig:critical}). In Fig.~\ref{fig:critical}
we also show the phase boundary as a function of $\delta_z/\pi$. The results compare very well 
with Eq.~(\ref{eq:alpha_c_spinless}).

\begin{figure}
\epsfxsize=6cm
\epsfbox{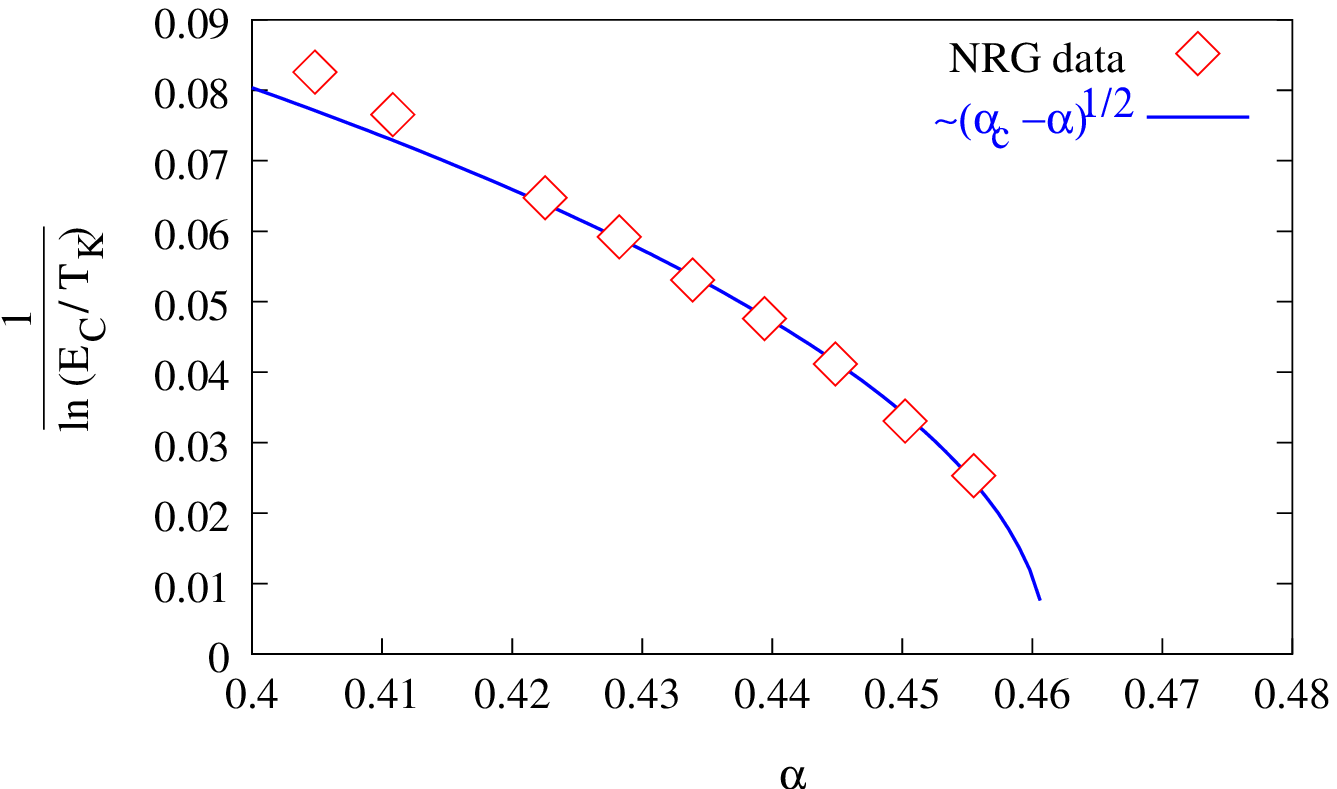}
\caption{(color online) Kondo temperature as a function of dissipation,
for $j_\perp=0.4$ and $j_z=0.1$} \label{fig:TK}
\end{figure}

The dissipation dependence of the Kondo 
temperature can be also easily extracted from  the finite size spectrum. 
In Fig.~\ref{fig:TK} we show  $1/{\rm ln}(E_C/T_K)$ as a function of 
$\alpha$. According to  Eq.(~\ref{eq:TK_nospin}) this should vanish as 
$\sim \sqrt{\alpha_c-\alpha}$ as one approaches the phase boundary,
implying  an extremely fast change in $T_K$ as one approaches  
$\alpha_c$. Though it is very difficult numerically to get to 
the real scaling regime, the obtained results are quite consistent with the 
analytical expression Eq.(~\ref{eq:TK_nospin}).

\begin{figure}
\epsfxsize=6cm
\epsfbox{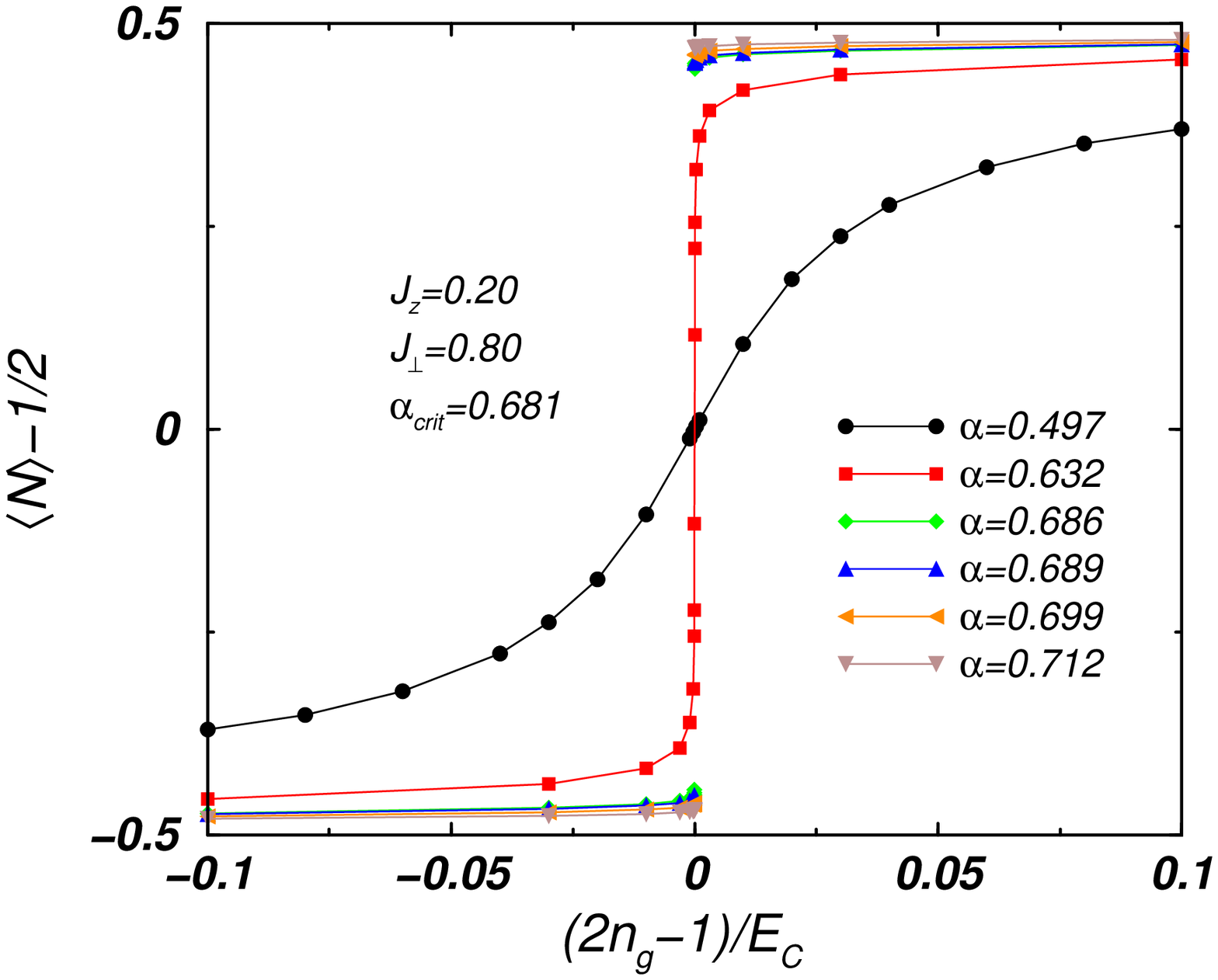}
\caption{(color online) The expectation value of  $\hat N - 1/2$ 
for $j_z=0.2$, $j_\perp=0.8$ and different values of
$\alpha$. For moderate values of the bosonic coupling the step is
smeared out by quantum fluctuations of the charge of the dot. 
The step gets sharper and sharper  as $\alpha$ approaches  $\alpha_c$, and
For $\alpha > \alpha_c$ a jump appears in $\langle \hat N \rangle$.} 
\label{fig:magnetization}
\end{figure}

To compute the full shape of the $\langle \hat N\rangle (n_g)$ curve, 
one has to study the model for gate voltages $n_g\ne1/2$, which corresponds 
to a field $B = (1-2n_g) E_C$ applied to the orbital spin $T_z= (1-\hat N)/2$ 
within our approach.
The result is shown in Fig.\ref{fig:magnetization}. Clearly, for $\alpha<\alpha_c$ 
 the Kondo correlations smear out the charging step of the grain. 

The Kondo temperature decreases as we increase the
coupling of the spin to the bosonic environment, and results in the sharpening of the
step in $\langle \hat N \rangle$. If we increase the bosonic coupling
further, the Kondo effect disappears and $\langle \hat N \rangle$
jumps at $n_g = 1/2$. It is difficult to tell from the 
numerics if the jump remains finite as $\alpha$ approaches 
$\alpha_c$ from above. 

However, since the phase transition 
is of Kosterlitz-Thouless type, one expects that the 'order parameter', 
{\em i.e.} the size of the jump approaches a constant as 
$\alpha$ approaches $\alpha_c$ from the dissipation-dominated (localized) 
side\cite{LeHur}.

The NRG method also enables us to compute local correlation functions, directly related 
to measurable quantities. To determine the scattering properties of the conduction electrons, 
we determined the conduction electrons' $T$-matrix. In the present case, the imaginary part 
of this  quantity is related to the renormalized conductance of the box-lead 
junction. The computation of this $T$-matrix can be reduced to the determination of 
the so-called 'composite Fermions' spectral function that can be computed by NRG 
similar to the single channel Kondo problem, \cite{costi2000,zarand2004}
\begin{equation}
{\rm Im}T_\mu(\omega)= A \;\sum\limits_n\left|\langle
n|F^\dagger_\mu|0\rangle\right|^2\delta(\omega-(E_n-E_0))\;,
\label{eq:T}
\end{equation}
where the composite fermion operator is given as
$$
F^\dagger_\mu=\sum\limits_{k\mu}\vec{S}\vec{\sigma}_{\mu\nu}c^\dagger_{k\nu}\;.
$$
The constant $A$ in Eq.~(\ref{eq:T}) can be determined from the condition, 
${\rm Im}T_\mu(\omega) = 2 {\rm sin}^2(\delta/\pi)$, where $\delta$ denotes the 
phase shift associated with the fixed point spectrum, and is $\delta=\pi/2$ at the 
quantum fluctuation dominated Kondo fixed point.

\begin{figure}
\epsfxsize=6cm
\epsfbox{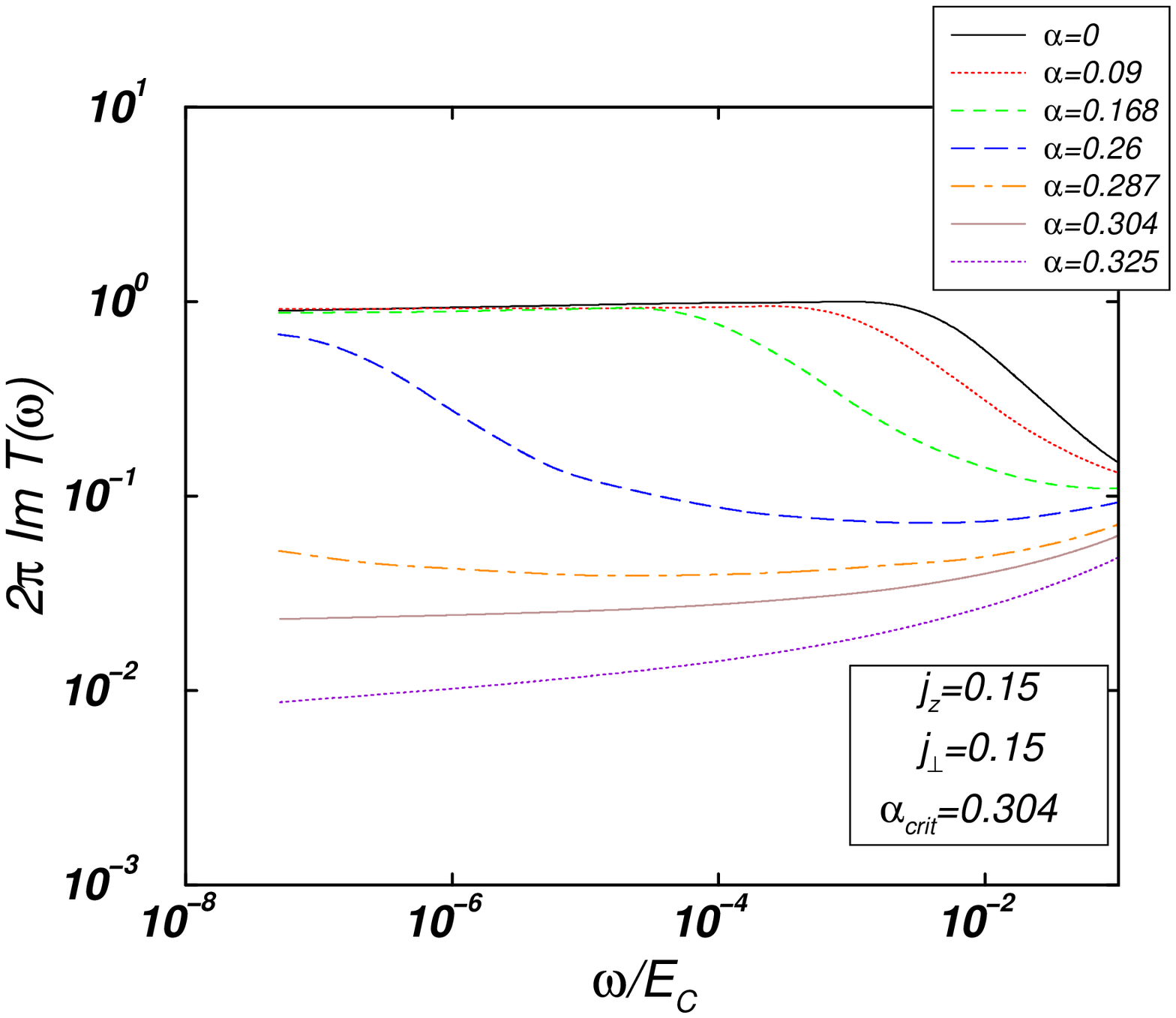}
\caption{(color online) The imaginary part of the fermions' $T$-matrix for
$j_z=0.15$, $j_\perp=0.15$ and for different values of the bosonic
coupling. For decoupled bosonic heat bath, the usual Kondo
resonance shows up. As $\alpha$ is increased, the width of the
Kondo resonance is decreasing and at the critical bosonic coupling
($\alpha_c=0.304$) it disappears and the electrons become decoupled
from the spin. } \label{fig:composite}
\end{figure}

The results are shown in Fig.\ref{fig:composite}. These results show that 
in the fermionic phase the dot-lead junction opens up, and the transmission between the 
box and the lead becomes perfect at $T=0$ temperature. 
It is adequate to mention here that the 
$T$-matrix computed here is much less accurate than that of
the one channel Kondo model. The reason is that due to the much
bigger Hilbert space we lose a substantial part of the spectral weight. 

\section{Conclusions}
\label{sec:conclusions} 

In the present paper we studied the dissipation-induced phase transition 
at the degeneracy point of a single electron box, discovered by Le Hur, 
using non-perturbative methods  such as bosonization and numerical renormalization 
group (NRG). Throughout this paper we focused to the limit where the level spacing of the 
box is negligible. Our calculations confirm the picture of Ref.~\onlinecite{LeHur}:
For small dissipation  the dissipative coupling is irrelevant
and charging steps of the isolated dot become rounded. However, above a critical 
value  the  dissipative coupling caused by fluctuations of the electromagnetic  field
dominates, and suppresses quantum fluctuations of the 
charge on the dot. As a results, the charging steps are restored for large 
enough dissipation. 

Our non-perturbative analysis reveals that 
- as already found by the perturbative analysis of Ref.~\onlinecite{LeHur} - 
this phase transition is of Kosterlitz-Thouless type. Consequently, 
the Kondo energy, where quantum fluctuations start to dominate 
goes to zero exponentially fast as $\alpha$ approaches its critical 
value $\alpha_c$.  

This phase transition also shows up in the {\em conductance} of the junction, 
which has a {\em jump}. This can be seen as follows. For $\alpha >\alpha_c$ 
the scaling equations imply that $j_\perp$ scales to 0. Since the linear conductance 
between the lead and the box is proportional to $j_\perp^2\sim t^2$, this implies that 
the liner conductance  for $\alpha>\alpha_c$ is identically zero. 
For $\alpha<\alpha_c$, on the other hand, the system flows to the Kondo (two-channel Kondo) fixed point, 
where the conductance is $g=1$  both in the spinless and in the spinful case.
As a result, the $T=0$ temperature conductance has a {\em jump}:
\be
g(T=0,\alpha) = {1\over 2} \theta(\alpha_c-\alpha)\;.
\ee

We can also determine the conductance at the critical value of $\alpha$ as a function of temperature
from our non-perturbative equations. 
From the scaling equations we obtain that at $\alpha=\alpha_c$ the coupling 
$j_\perp$ scales to 
zero approximately as 
\be
j_\perp(T) \approx {1\over {\rm ln} (T^*/T)}\;, \phantom{nnn}(\alpha=\alpha_c)
\ee
where $T^*$ is a dynamical scale. As a result, the box-lead 
conductance scales to zero as
\be
g(T) \sim {1\over {\rm ln}^2\;T^*/T}\;. \phantom{nnn}(\alpha=\alpha_c)
\ee

The NRG method makes us possible to determine the phase boundary for the case of spinless Fermions
between the  two phases non-perturbatively. For small values of the tunneling we find
from the scaling analysis that
\be 
\alpha_c \approx j_\perp, 
\ee
which agrees very well with the NRG results. This condition translates to the following 
condition for the device
\be
{C_g^2\over 4 C^2} {R_g\over R_Q}\approx {2\over \pi} \sqrt{g} \;,
\ee
where $g$ is the dimensionless conductance of the junction, $C_g$ is the gate capacitance, 
$C$ is the total capacitance of the island, and $R_g$ is the resistivity of 
the gate electrode. In other words, the above phase transition can be reached by 
increasing the resistance of the gate electrode or by changing the box-lead conductance.

Let us finally make a remark on the finite level spacing $\Delta$. All considerations
above hold only in the limit of vanishing level spacing. In reality, however, this level 
spacing is {\em finite}, and plays a role similar to  a finite temperature. As a 
consequence, quantum fluctuations remain always finite at the degeneracy point, 
and it is not clear if a real phase transition occurs even in the presence of 
 a finite level spacing.

We would like to thank K. Le Hur for valuable discussions. 
This research has been supported by
NSF-MTA-OTKA Grant No. INT-0130446, Hungarian Grants No. OTKA
T038162, T046267, D048665, and T046303, and the European 'Spintronics' RTN
HPRN-CT-2002-00302.


\begin{thebibliography}{99}
\bibitem{note1} The value of $n_g$ where the number of electrons changes
is slightly shifted with respect to  half-integers due to the finite level spacing 
on the box.
\end{thebibliography}
\end{document}